\documentclass[prc,aps,a4paper,groupedaddress,superscriptaddress,nofootinbib,showpacs
,preprintnumbers,onecolumn]{revtex4}
\usepackage{graphicx}
\usepackage{amsfonts}
\usepackage{amssymb,ulem}
\usepackage{amsmath}
\usepackage{natbib}
\usepackage{dcolumn}
\usepackage{bm} 
\newcommand{\bwt}{\begin{widetext}}
\newcommand{\ewt}{\end{widetext}}
\newcommand{\beq}{\begin{equation}}
\newcommand{\eeq}{\end{equation}}
\newcommand{\bea}{\begin{eqnarray}}
\newcommand{\eea}{\end{eqnarray}}
\begin{document}
\title{Quark matter under strong magnetic field in  chiral models.}
\author{Aziz Rabhi}
\email{rabhi@teor.fis.uc.pt}
\affiliation{Centro de F\' {\i}sica Computacional, Department of Physics, University of Coimbra, 3004-516 Coimbra, Portugal} 
\affiliation{Laboratoire de Physique de la Mati\`ere Condens\'ee,
Facult\'e des Sciences de Tunis, Campus Universitaire, Le Belv\'ed\`ere-1060, Tunisia}
\author{Constan\c ca Provid\^encia}
\email{cp@teor.fis.uc.pt}
\affiliation{Centro de F\' {\i}sica Computacional, Department of Physics, University of Coimbra, 3004-516 Coimbra, Portugal}
\date{\today}    
\begin{abstract}
The chiral model is used to describe quark matter under strong magnetic fields and
compared to other models, the MIT bag model and the two flavor Nambu-Jona-Lasinio model. The
effect of vacuum corrections due to the magnetic field is discussed. It is shown that if the
magnetic field vacuum corrections are not taken into account explicitly  the
parameters of the models should be fitted to low density meson properties in the presence of
the magnetic field.  
\end{abstract}
\maketitle

\section{Introduction}

Magnetars, neutron stars with very strong magnetic fields of the order of
$10^{14}-10^{15}$ G at the surface, are  sources of very energetic electromagnetic 
radiation, mainly  gamma and X rays \cite{duncan,usov,pacz}. Presently, about 21 of 
these objects have been detected, most of them as
soft gamma repeaters (SGRs) and anomalous X-ray pulsars (AXPs) \cite{index}.

It has been argued \cite{bodmer} that {\it strange quark matter} (SQM), {\it i.e.} 
quark matter with strangeness per baryon of the order of unity, may be the true
ground state of hadronic matter. This could imply that compact stars are mainly quark 
stars (see also \cite{bombaci}). Magnetars as compact quark stars have been first investigated
in \cite{chakrabarty96}, where the MIT bag model \cite{bag} was applied to obtain the equation of state (EOS) of stellar quark matter under magnetic fields as strong as $10^{18}$ G. The EOS for magnetized
quark stars  described within the MIT bag model and taking into account the anomalous magnetic
moment of quarks (AMM) was studied in \cite{aurora}.
The su(2) version of the Nambu--Jona--Lasinio model (NJL) \cite{njl}, an
effective  model which includes the  chiral symmetry, was applied to study the 
stability of quark matter under very strong magnetic fields in \cite{klimenko03}. The same
model and its su(3) extension were used to describe quark stars with very large magnetic
fields in \cite{njlsu2,njlsu3}. 
In these papers, the phenomenon of magnetic catalysis within the NJL models  has
been discussed.
 Magnetic catalysis
is one of the most important effects of the magnetic field in quark models with
chiral symmetry and  corresponds to the  enhancement of the chiral symmetry breaking in
the magnetic field \cite{catalysis}. Another  nontrivial
effect of the magnetic field  is the possibility that
strong magnetic fields can turn a crossover into a first order QCD
transition \cite{fraga08}.

A different quark model with chiral symmetry, the chiral model of pions and quarks or 
nucleons~\cite{gell60,KRS84, kbk90}, also known as linear sigma model, was applied to study the high density npe$\mu$ matter with
$\pi^0$-condensation~\cite{takahashi2006}. In the present work we will use the same model
to describe both symmetric quark matter and stellar quark matter under strong magnetic
fields. No pion condensation will be considered. The results will be compared with the MIT and
NJL models. In particular we will investigate the inclusion of vacuum corrections due to the
strong magnetic field.

The linear sigma model coupled to quarks and the Polyakov loop has been recently used to 
study the phase diagram of hot QCD in a strong external magnetic field \cite{mizher10}. There
it was shown  that a strong  magnetic field could give rise to a splitting of the deconfinement
and chiral transitions if the   B dependent vacuum corrections were included. These results 
agree well with the diagram coming
from the NJL model \cite{gatto10}, where the vacuum corrections are present
authomatically, and with the results of the
lattice calculations \cite{elia10}.

The MIT bag model describes quarks as a free gas of quarks already in a chiral restored
state. The bag pressure provides the confinement and is just a parameter which can be fixed
from the nucleon sector. The problem of chiral symmetry restoration is beyond the scope of this
model. Both NJL and the chiral model are described by chiral symmetric Lagrangian densities and
a vacuum state with spontaneously broken chiral symmetry. In the chiral model, the chiral condensate plays the role of the  bag pressure and its value in the vacuum is fixed from the pion decay constant, which is well known, and the sigma mass. The connection between the MIT bag model and the chiral model has been discussed in~\cite{SB04}. In the NJL model the model parameters are fixed by fitting the pion decay constant and the quark condensate.

In section II we make a brief review of the three models and their corresponding EOS under the effect of
a magnetic field and discuss how the  model parameters are fixed for a finite magnetic field. Results are discussed in section III and conclusions are drawn in section IV.

\section{Quark models}
In the present section we give a brief review of the quark models including the effect of a
strong magnetic field used in this study: the chiral sigma model, the su(2) NJL model and the
MIT bag model. 

\subsection{Chiral sigma model}
 
We consider the chiral sigma model for quarks interacting with an external magnetic field. The chiral symmetric Lagrangian density reads \cite{gell60, kbk90} 

\bea
{\cal L}&=& \bar{\psi}_{f}\left(i\gamma_{\mu}\partial^{\mu}-\hat q\gamma_{\mu}A^{\mu}
-g\left(\sigma +i\gamma_{5}\vec{\tau}.\vec{\pi}\right) \right )\psi_{f} \cr
&+&\frac{1}{2}\partial_{\mu}\sigma \partial^{\mu}\sigma
+\frac{1}{2}\partial_{\mu}{\pi}^0\partial^{\mu} {\pi}^0+ D_{\mu}{\pi}^+ D^{\mu} {\pi}^-
-U\left(\sigma, \vec{\pi}\right)-\frac{1}{4} F^{\mu \nu}F_{\mu \nu},
\label{lagran}
\eea
where $\psi_{f}$ is the quark field, $\sigma$ and ${\pi}^0, \, \pi^\pm=(\pi^1\pm i\pi^2)/\sqrt{2}$ are the meson fields, 
$D^\mu=\partial^\mu+i e A^\mu$,
$g$ is the
quark-meson coupling constant, $\hat q=(\frac{1}{3}+\tau_3)e/2$ and $A^\mu=(0,0,Bx,0)$ 
refers to an external magnetic field along the z-axis.

 The potential functional $U$ 
is a ''Mexican hat'' potential, which leads to spontaneous chiral symmetry breaking and is
included to reproduce the vacuum expectation value of the sigma field. For  exact chiral symmetry ($m_\pi=0$), the potential is 
\beq
U\left(\sigma, \vec{\pi}\right)=\frac{\lambda^2}{4}\left(\sigma^2+\vec{\pi}^2-f^2_{\pi} \right)^2.
\label{pot} 
\eeq
where $\displaystyle\lambda^2=\frac{m^{2}_{\sigma}}{2 f^2_{\pi}}$, and $m_{\sigma}$ is the mass of the $\sigma$-meson.
The vacuum expectation value of the field $\sigma$ is $\left\langle \sigma\right\rangle
=f_{\pi}$, where $f_{\pi}=93$ MeV is the pion decay constant in the absence of a magnetic field. 
In the normal phase with no pion condensation, the energy density is given by
\beq
{\cal E}=\sum_{f=u,d}\frac{N_c
|q_{f}|B}{2\pi^{2}}\sum^{\nu_{f,max}}_{\nu=0}\alpha_\nu\int^{p_{F, f\nu}}_{0}dp_{z}\sqrt{p_{z}^{2}+m^{2}+2\nu |q_{f}|B} 
+ U\left(\sigma, 0\right)
\eeq
where $\displaystyle m=g\sigma$, and $\displaystyle\nu=n+\frac{1}{2}-sgn(q)\frac{s}{2}=0, 1, 2,
\ldots$ enumerates the Landau levels (LL) of the fermions with electric charge $q$, the factor
$\alpha_\nu=1,\, (2)$ for $\nu=0$ ($\nu\ge 1$)
 takes care of singly degenerate zeroth LL and doubly degenerate LL levels with $\nu\ge 1$ and 
the Fermi momentum of LL $\nu$ is $p_{F,f\nu}= \sqrt{\mu^2_f-M^2_{f}({\nu},B)}$. The coefficient $N_c=3$ stands for the color degeneracy. We use $m_{\sigma}=1200$MeV.

The energy minimization for each baryon density with respect to the $\sigma$ field,  gives the so called gap equation
\beq
\sum_{f=u, d}\frac{N_c |q_{f}|B}{2\pi^{2}}
\sum_{\nu}^{\nu_{f,max}}\alpha_\nu\int^{p_{F, f
\nu}}_{0}\frac{dp_{z}}{\sqrt{p_{z}^{2}+m^{2}+2\nu |q_{f}|B}}
+\frac{\partial U}{\partial\sigma}=0
\eeq
The mass in the vacuum is  $\displaystyle m_0=g f_{\pi}$. In the presence of a strong magnetic
field the vacuum properties are affected and a quark vacuum correction  should be included
\cite{njlsu2,mizher10}. This correction was not considered in~\cite{takahashi2006}. Since the magnetic
fields discussed were below $6\times 10^{18}$ G, we do not expect a large effect coming from
this term.  We will take into account this correction in two different ways: a)  we will redefine the constant $f_\pi$ and
suppose all vacuum effects are described by  the ``Mexican hat'', including the magnetic field
contribution; b) we consider that the ``Mexican hat'' potential does not include the magnetic
field vacuum contribution and we will add this contribution explicitly as  an extra term to
(\ref{pot})  just like it  was done in \cite{mizher10}.

\subsection{The MIT bag model}
 In the presence of a strong magnetic field,  the
energy density 
and quark density within the MIT bag model, are given by \cite{aziz09}
\bwt
\bea
\varepsilon_m &=& \sum_{f=u,d}
\frac{N_c|q_{f}|B}{4\pi^2}
\sum^{\nu_{f,max}}_{\nu=0}\alpha_\nu \left[\mu_f\, p_{F,f\nu}
+M^2_{f}({\nu},B)\ln\left|\frac{\mu_f+p_{F,f\nu}}{M_f({\nu},B)}\right|  \right]
+\hbox{Bag}, \label{ener}\\
\rho_q &=&
 \sum_{f=u,d}\frac{N_c|q_{f}|B}{2\pi^2}\sum^{\nu_{f,max}}_{\nu=0}\alpha_{\nu}p_{F,f\nu}
\label{dens}
\eea
\ewt
where $M_f({\nu},B)=\sqrt{m^2_f+2\nu |q_f| B}$ and $\nu $ runs over the allowed
LL,  
$m_q$ is  the quark mass and Bag represents the bag pressure. We only consider flavors $u$ and $d$.

\subsection{The su(2) NJL model}

We  consider the two flavor NJL model  defined by  the following Lagrangian density \cite{njlsu2,njlsu3}
\begin{equation}
{\cal L} = {\cal L}_{f} - \frac {1}{4}F_{\mu
\nu}F^{\mu \nu}
\end{equation}
where the quark sector is described by the  Nambu--Jona-Lasinio model
\begin{eqnarray}
{\cal L}_f &=& {\bar{\psi}}_f \left[\gamma_\mu\left(i\partial^{\mu}
- \hat q A^{\mu} \right)-
m_c \right ] \psi_f \nonumber \\
&+& G \left [({\bar \psi}_f \psi_f)^2 + ({\bar \psi}_f i\gamma_5
{\vec \tau} \psi_f)^2 \right ]\;, \label{njl}
\end{eqnarray}
$m_c = m_u \simeq m_d$ are the quark  current masses;  $A_\mu$ and $F_{\mu \nu }=\partial
_{\mu }A_{\nu }-\partial _{\nu }A_{\mu }$ are used to account
for the external magnetic field. Since we are interested in a static and constant magnetic field
in the $z$ direction, $A_\mu=\delta_{\mu 2} x_1 B$.

The energy density is given by
\begin{equation}
{\cal E} (\mu_f,B)= -P^N + \sum_{f} \mu_f \rho_f \,\,\,,
\end{equation}
 where $\mu_f$ is the chemical potential of flavor $f$ and the pressure is $P^N = P(\mu_f)|_{M(\mu_f)} - P(0)|_{M(0)}$ with
\begin{equation}
P = \theta_u+\theta_d  -G(\phi_u+\phi_d)^2\, ,
\end{equation}
For a given flavor, the  $\theta_f$ term is given by
\begin{equation}
 \theta_f=-\frac{i}{2}  {\rm tr}  \int  \frac {d^4 p}{(2\pi)^4} \ln \left(-p^2 + M^2 \right )
\end{equation}
and the condensate  $\phi_f= \langle {\bar \psi}_f \psi_f \rangle$, so that \cite{njlsu2,njlsu3}
 \begin{equation}
P= \left (P^{vac}+P^{mag} + P^{med} \right )\,\,,
\label{pressBmu2}
\end{equation}
where the vacuum contribution reads
\begin{equation}
P^{vac}=- \frac{N_c \, N_f}{8\pi^2} \left[ M^4 \ln \left(
    \frac{\Lambda+ \epsilon_\Lambda}{M} \right )
 - \epsilon_\Lambda \, \Lambda\left(\Lambda^2 +  \epsilon_\Lambda^2 \right ) \right],
\end{equation}
with  $\epsilon_\Lambda=\sqrt{\Lambda^2 + M^2}$, $\Lambda$ representing a non covariant ultra
violet cut off;   the  finite  magnetic contribution is
\begin{equation}
P^{mag}= \sum_{f=u,d}\frac {N_c (|q_f| B)^2}{2 \pi^2} \left [ \zeta^\prime(-1,x_f) -  \frac {1}{2}( x_f^2 - x_f) \ln x_f +\frac {x_f^2}{4} \right ]\,\,,
\label{pmag0}
\end{equation}
with   $x_f = M^2/(2 |q_f| B)$ and 
$\zeta^\prime(-1,x_f)= d \zeta(z,x_f)/dz|_{z=-1}$ where $\zeta(z,x_f)$ is the Riemann-Hurwitz
zeta function, and the medium contribution can be written as
\begin{eqnarray}
P^{med}_{M}&=&\sum_{f=u,d}\frac {|q_f| B N_c }{4 \pi^2}\sum_{\nu=0}^{\nu_{f,max}} \alpha_\nu
\left [ \mu_f \, p_{F,f\nu}
- M _f(\nu,B)^2 \ln \left| \frac { \mu_f + p_{F,f\nu}} {M_f(\nu,B)} \right| \right ] - 2G(\phi_u+\phi_d)^2 ,
\label{PmuB}
\end{eqnarray}
where  $M_f(\nu,B)
= \sqrt {M^2 + 2 |q_f| B \nu}$ and  $p_{F,f\nu}=\sqrt{\mu_f^2 - M_f(\nu,B)^2}$.
The  upper Landau level (or the nearest integer) is defined by
\begin{equation}
\nu_{f, max} = \frac {\mu_f^2 -M^2}{2 |q_f|B}= \frac{p_{f,F}^2}{2|q_f|B}.
\label{landaulevels}
\end{equation}
The effective quark masses can be obtained self consistently  from
\begin{equation}
 M=m_c - 2 G (\phi_u+\phi_d),
 \label{mas}
\end{equation}
where the condensates $\phi_f$ are given by
\begin{equation}
\phi_f=(\phi_f^{vac}+\phi_f^{mag}+\phi_f^{med})_{M}
\end{equation}
with
\begin{eqnarray}
\phi_f^{vac} &=& -\frac{ M N_c }{2\pi^2} \left [
\Lambda \epsilon_\Lambda -
 {M^2}
\ln \left ( \frac{\Lambda+ \epsilon_\Lambda}{{M }} \right ) \right ]\,\,,
\end{eqnarray}
\begin{eqnarray}
\phi_f^{mag}
&=& -\frac{ M |q_f| B N_c }{2\pi^2}\left [ \ln \Gamma(x_f) -\frac {1}{2} \ln (2\pi) +x_f -\frac{1}{2} \left ( 2 x_f-1 \right )\ln (x_f) \right ] \,\,,
\end{eqnarray}
and
\begin{eqnarray}
\phi_f^{med}&=&\frac{ M |q_f| B N_c }{2 \pi^2}
\sum_{\nu=0}^{\nu_{f,max}} \alpha_\nu 
\ln \left ( \frac { \mu_ f +p_{F,f\nu}} {M_f(\nu,B)} \right )\,\,,
\label{MmuB}
\end{eqnarray}
The density, $\rho_f$, corresponding to each different flavor,  is given by Eq.~(\ref{dens}).

For the su(2) NJL model we use the same parametrization as given in~\cite{buballa96}: $\Lambda=587.9$ MeV, $m_{c}=5.6$ MeV, $m_e=0.511$ MeV, $m_\mu=105.66$ MeV, and $G\Lambda^2=2.44$ which gives a quark vacuum mass equal to 400 MeV in the absence of a magnetic field.

\subsection{Fixing the model parameters at finite B}
\label{sec:para}

\begin{figure}[ht]
\vspace{1.5cm}
\centering
\includegraphics[width=0.5\linewidth,angle=0]{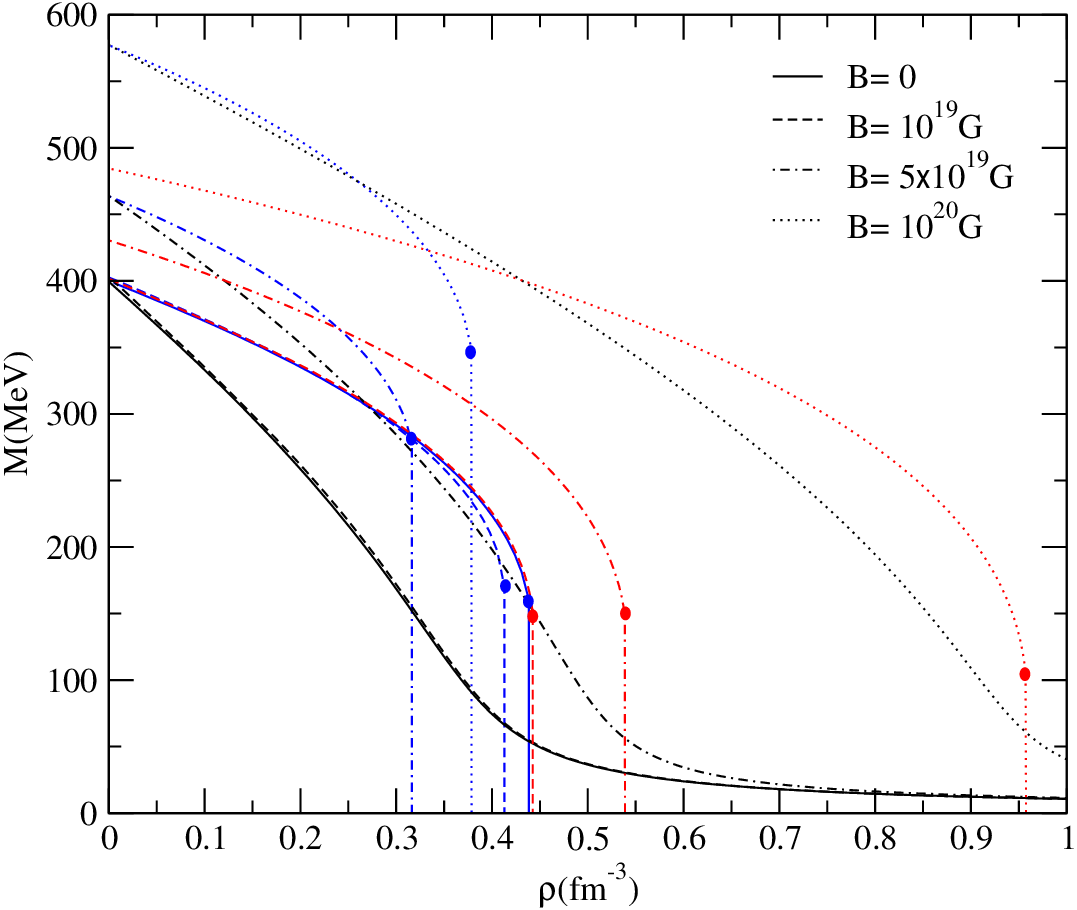}
\caption{(Color online) The quark mass  as a function of baryon number density for isospin-symmetric quark
matter, and for several values of the magnetic field. The lines
correspond to the su(2) NJL (black lines),  the chiral model I with a renormalized $f_\pi$
(blue lines) and the chiral model II with $B$ dependent vacuum corrections defined in
(\ref{pmag}) (red lines).}
\label{mass}
\end{figure}

In order to compare the MIT bag model with the other two models we chose
 $\hbox{Bag}=\, 0.9\, \hbox{fm}^{-4}=\, 177.3\, \hbox{MeV}/\hbox{fm}^{3}$ which is in between the values we get for the
chiral model and the NJL model in the chiral symmetric density region. We note that in MFT,
when the chiral symmetry is restored, \textit{i.e.}($\left\langle \sigma\right\rangle  =0 $ and
$\displaystyle\left\langle \vec{\pi}\right\rangle=0 $), the potential functional reduces to a
constant energy density equal to $\displaystyle\frac{\lambda^2}{4}f^{4}_{\pi}$, and the
constituent quark mass vanishes, leaving free massless quarks. 
The chiral Lagrangian, in the region where the chiral symmetry is restored, can be identified by a MIT bag quark model with a bag pressure Bag$\displaystyle=\frac{\lambda^2}{4}f^{4}_{\pi}$.

In order to be able to establish a comparison between the chiral model and the NJL model, 
we include, in the  chiral model, vacuum corrections due to the magnetic field in
two different ways: 
(a) by  renormalizing the pion constant $f'_\pi$ constant so that the vacuum  quark mass
$m_0=g \,  f'_\pi$ in the chiral model coincides with the one obtained with the NJL model. We
will call this model the chiral model I; 
(b) by including in the potential (\ref{pot}) the explicit field vacuum contribution,
$U_{mag}(\sigma,B)=- P^{mag}$, where $P^{mag}$ was defined in (\ref{pmag0}) with $M=g\sigma$ as
was done in \cite{mizher10}. This will be the chiral model II.
In this case we have for the potential
\beq
U\left(\sigma, \vec{\pi},B\right)=\frac{\lambda^2}{4}\left(\sigma^2+\vec{\pi}^2-f^2_{\pi} \right)^2+U_{mag}(\sigma,B)-U(\sigma_0,0,B)
\label{potb}
\eeq
where
\begin{equation}
U_{mag}(\sigma,B)= \sum_{f=u,d}\frac {N_c (|q_f| B)^2}{2 \pi^2} \left [ \zeta^\prime(-1,x_f) -  \frac {1}{2}( x_f^2 - x_f) \ln x_f +\frac {x_f^2}{4} \right ]\,\,,
\label{pmag}
\end{equation}
with  $x_f = (g\sigma)^2/(2 |q_f| B)$ as in  (\ref{pmag0}), and $\sigma_0$ is the $\sigma$ field in the vacuum. The last term in (\ref{potb}) insures that the pressure goes to zero at zero density.
Replacing this expression for the potential in the gap equation we get:
\bea
\sum_{f=u, d}\frac{N_c |q_{f}|B}{2\pi^{2}}
&&\sum_{\nu}^{\nu_{f,max}}\alpha_\nu\int^{p_{F, f
\nu}}_{0}\frac{dp_{z}}{\sqrt{p_{z}^{2}+m^{2}+2\nu |q_{f}|B}}
 +\frac{\lambda^2}{g m}\left(\sigma^2-f^2_{\pi} \right) \sigma\nonumber\\
&& -\frac{ |q_f| B N_c }{2\pi^2}\left [ \ln \Gamma(x_f) -\frac {1}{2} \ln (2\pi) +x_f -\frac{1}{2} \left ( 2 x_f-1 \right )\ln (x_f) \right ] 
=0
\eea

In Fig.~\ref{mass} we compare the quark masses in isospin-symmetric quark matter as a function of
the baryon density obtained with both models and using the two approaches described above to
take into  account the
B dependent vacuum corrections. By construction the curves for the NJL and the chiral model I
start at the same  mass at zero density and decrease with density reaching the value of
the current mass of the quarks in the chiral symmetric phase.
For the chiral model II 
the quark  vaccum mass also increases with B but not so strongly. 
 As discussed in~\cite{klimenko03,njlsu2,njlsu3}, 
in the NJL model, the magnetic field shifts the chiral symmetry restoration to larger
densities. The same occurs with the chiral model II. We also note that within this model the chiral symmetry restoration occurs slower than
in the NJL model, a feature that is already present for zero magnetic field.
In the chiral model I   magnetic catalysis is not so clear.
 There are two competing effects which can be identified from the gap equation: for a larger $f_\pi$, {\it i.e.} larger
vacuum quark mass, the restoration of
chiral symmetry occurs at smaller densities if no Landau quantization is present in the
quark quasi-particle (QP) energy. This is the effect that is observed for the smaller magnetic
fields. However, Landau quantization reduces the QP energy, and, therefore, for very strong
fields, the chiral symmetry restoration occurs at larger densities when B increases. 

\section{Results}

\begin{figure}[ht]
\vspace{1.5cm}
\centering
\includegraphics[width=0.85\linewidth,angle=0]{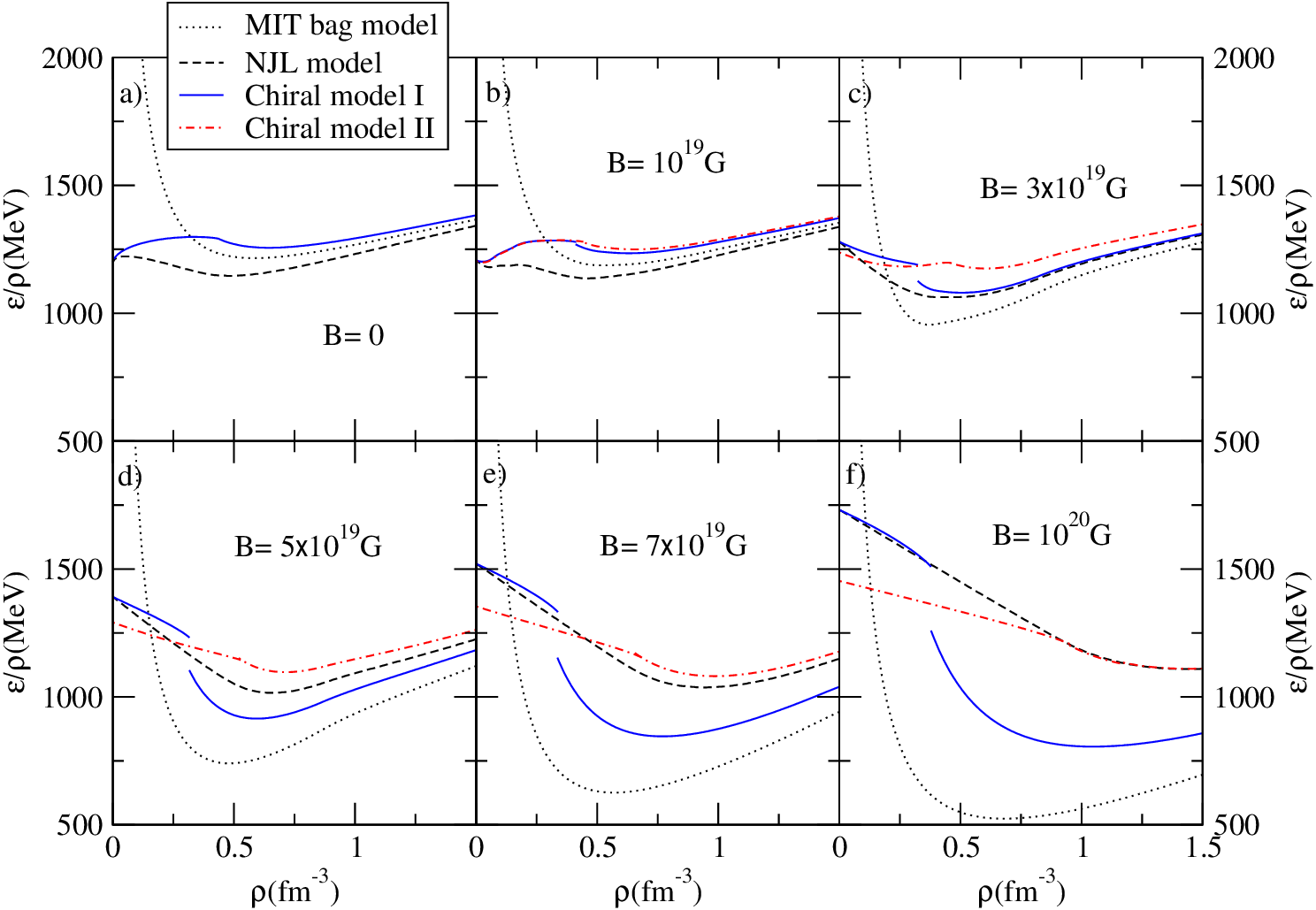}
\caption{(Color online) Quark matter energy per baryon number  as a function of the baryon number density for
isospin-symmetric quark matter, and for several values of the magnetic field. The different
lines correspond to the NJL (dashed lines), chiral model I (full lines), chiral model II (dash-dotted lines)
and MIT bag model (dotted lines). }
\label{bener}
\end{figure}

The main objective of the present section is to compare the different quark models and to discuss
how the vacuum corrections due to the presence of a strong magnetic field may be taken into
account in the  MIT bag model and in the chiral model. We will first compare the properties of
symmetric baryonic matter and in a second subsection, we will discuss the implications in the EOS
of stellar quark matter obtained with both chiral models considered.

\subsection{Symmetric quark matter}
\begin{figure}[ht]
\vspace{1.5cm}
\centering
\includegraphics[width=0.85\linewidth,angle=0]{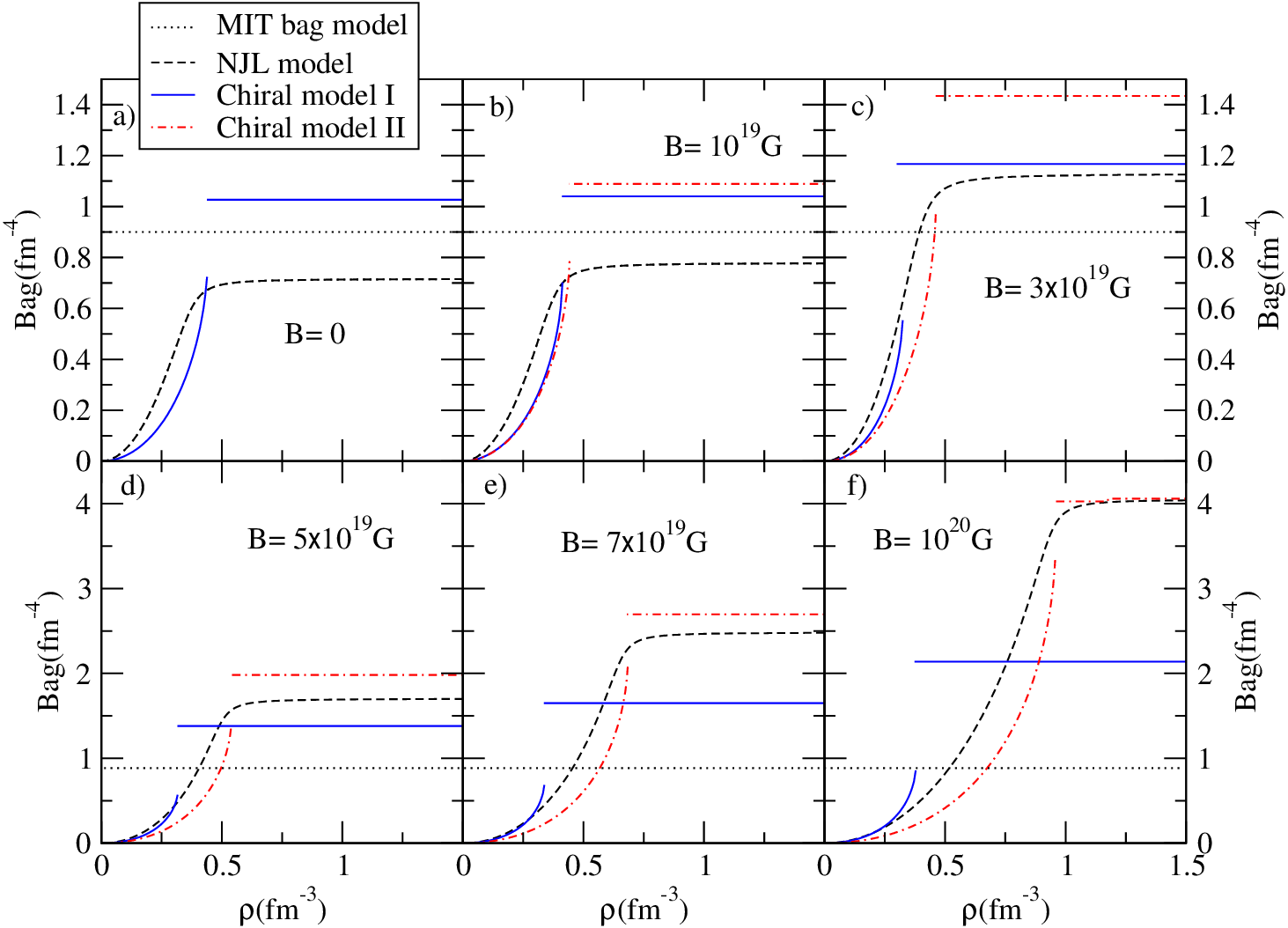}
\caption{(Color online) The bag pressure as a function of baryon number density for isospin-symmetric quark
matter.  The different
lines correspond to the NJL (dashed lines), chiral model I (full lines), chiral model II (dash-dotted lines)
and MIT bag model (dotted lines). Please notice that the $y$-axis scale of the bottom figures is different from the one of
the top figures.}
\label{bag}
\end{figure}

In Fig.~\ref{bener} the energy per baryon calculated with the chiral model, NJL model, and MIT bag model, 
is shown for several values of the magnetic field.
In the absence of an external magnetic field the three models give similar results at high
densities, when a chiral symmetry restored state is the ground state of the system. Since, both the
chiral model and the NJL model use low density meson properties to fix the parameters of the
models they behave in a similar way. Below $B=10^{19}$ G the magnetic field has no noticeable
effect on the energy per particle. However, above $B=10^{19}$ G the effect becomes stronger
and the MIT gives the lowest energy per particle. This is due to the fact that
the bag pressure, which describes the vacuum effects, was kept constant, independent of the
magnetic field, and, due to the Landau quantization, the contribution of the kinetic energy is
strongly reduced. The minimum of the energy per particle is shifted to higher densities but not
so strongly as in the other models.  In the chiral model I  we see that:
 a) in the  chiral symmetry broken phase the chiral model has a behavior
similar to the NJL for large values of B, and at  zero density the energy per
particle of both models always coincide, b) although B dependent vaccum corrections are partially taken into account and the
energy per particle does not decrease so much as in the MIT bag model, in the chiral restored
phase the energy per particle is still much smaller than the predictions of both the NJL and the chiral model with the vacuum correction
(\ref{pmag}) for $B> 5\times 10^{19}$ G.  In chiral model II, where the $B$ depedent vacuum corrections are properly taken into account in the
chiral model we
conclude that: a) in the chiral symmetry broken phase the chiral model predicts 
for $B<3\times 10^{19}$ G a larger  energy per particle than NJL. However, for  larger
values of $B$ the opposite occurs because the vacuum mass in this model does not increase so
much with $B$ than in the NJL model; b) in the chiral restored phase the energy per particle is
larger in the chiral model until a very large magnetic field ($10^{20}$ G).

\begin{figure}[ht]
\vspace{1.5cm}
\centering
\includegraphics[width=0.85\linewidth,angle=0]{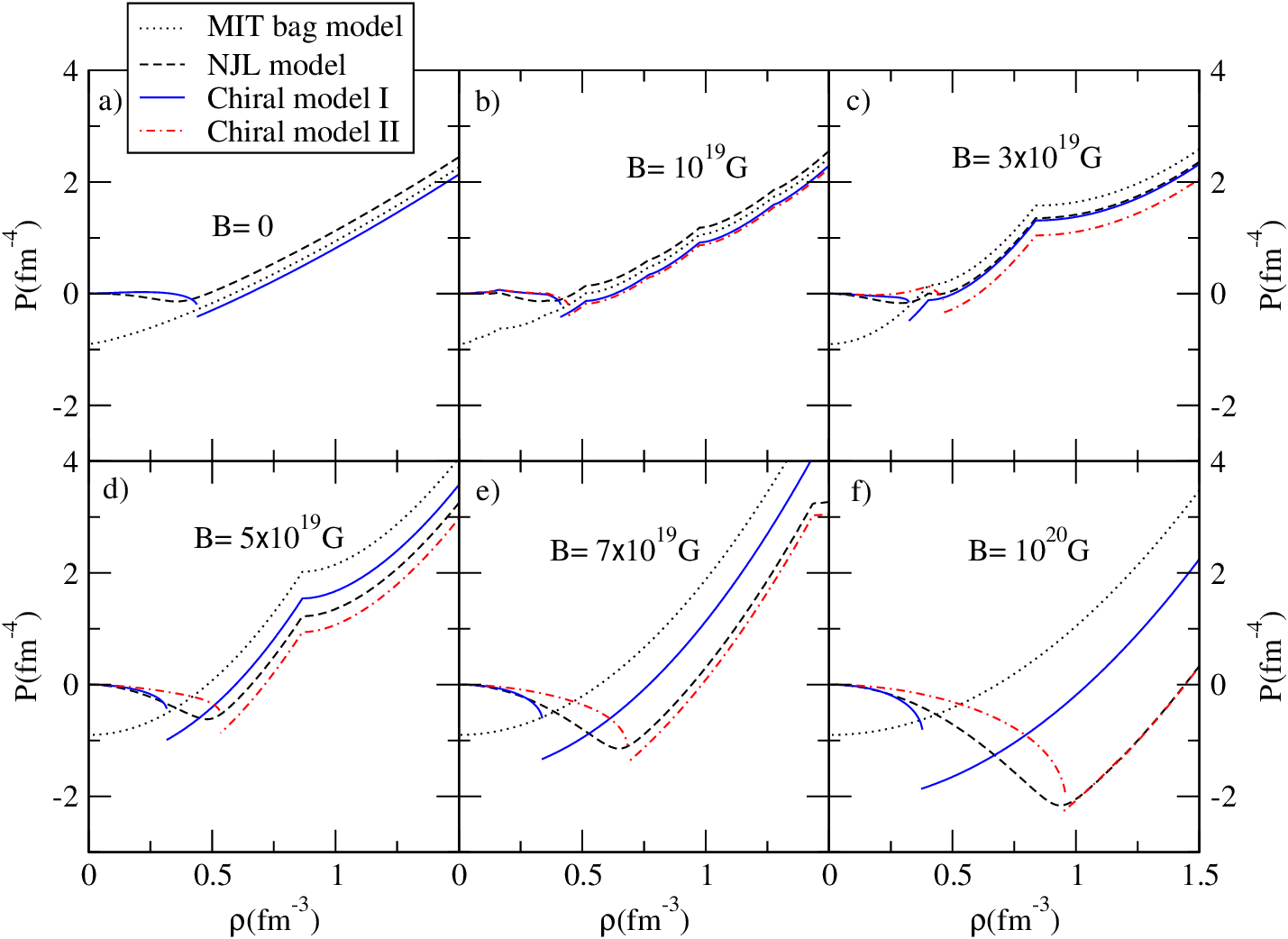}
\caption{(Color online) Pressure as a function of baryon number density for isospin-symmetric quark matter,
and for several values of the magnetic field. The different
lines correspond to the NJL (dashed lines), chiral model I (full lines), chiral model II (dash-dotted lines)
and MIT bag model (dotted lines).}
\label{press}
\end{figure}

Let us define, in all models, an effective bag pressure  which corresponds to the total energy minus the
kinetic energy contribution. In Fig.~\ref{bag} we plot the effective  bag pressure for the three
quark models,  and several magnetic field intensities. We have considered both approachs to the
B dependent vacuum corrections in the chiral model.
For $B=0$, all models are supposed to describe the same physics at high densities which
corresponds to  the chiral restored phase: the  NJL has the smallest bag pressure while 
the chiral model the largest one. For the MIT, as discussed before, we have chosen 
an intermediate value. For finite values of $B$, 
we conclude that: a) in the chiral symmetry broken phase the chiral model II has the smallest effective bag pressure although close to the one of the NJL model. The
chiral model I gets closer to the NJL model as  $B$
increases, however, the chiral symmetry restoration occurs at too low densities compared with
the other two chiral  models; b) in the chiral symmetry restored phase the effective bag
increases with $B$ in all models except in the MIT bag model. Within the NJL, the effective bag
increases faster with $B$ than in the chiral model II and at $B=10^{20}$ G both models
coincide. In the chiral model I the effective bag increases too slowly and above $4\times 10^{19}$ G it
is already smaller than the one obtained in the NJL model. However, we should point out that the large
differences occur only for very intense magnetic fields, above $5\times 10^{19}$ G.

The behavior of the bag pressure in the different models reflects itself on the pressure of the
system. In Fig.~\ref{press} we plot the pressure of the gas of symmetric quark matter as a
function of density for different magnetic field intensities. Below $10^{19}$ G all
models give results in the chiral restored phase similar to the ones obtained at $B=0$: the
model with the smallest effective bag pressure (the NJL model) has the largest pressure.
Above  $3\times 10^{19}$ G the different
effective bag pressures of the three models give rise to quite different pressures at high
densities with the MIT bag model predicting the largest one and the chiral model II the smallest one.

\subsection{Stellar quark matter}
\begin{figure}[ht]
\vspace{1.5cm}
\centering
\includegraphics[width=0.85\linewidth,angle=0]{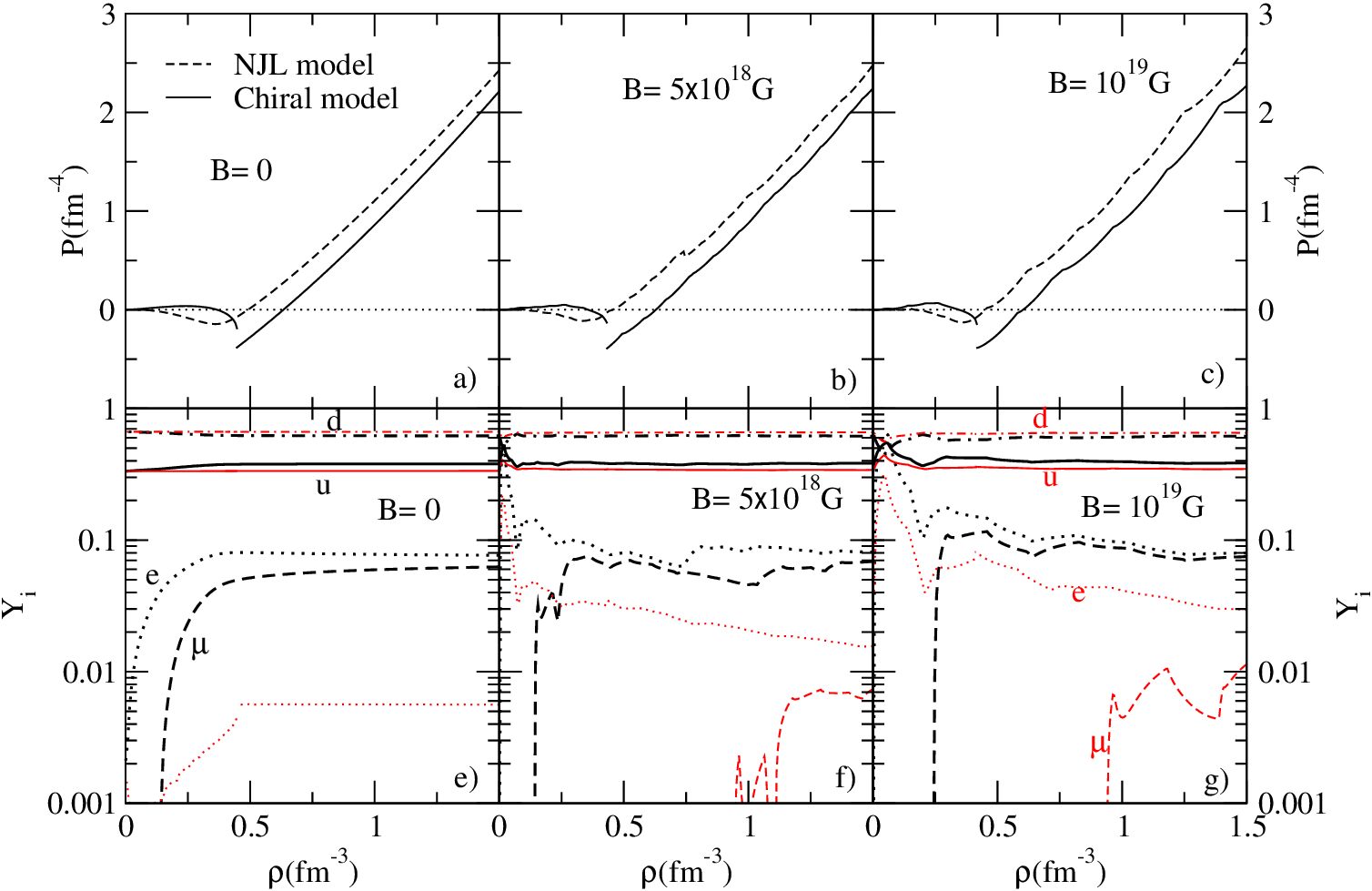}
\caption{(Color online) Pressure (top) and particle fractions (bottom) as a function of baryon number density for asymmetric quark matter,
and for  $B=0,\, 5\times 10^{18},\, 10^{19}$ G. The thin line is for the su(2) NJL model and
the thick line is for
the chiral model. The zero-axis is shown in the pressure plots with a thin dotted line. In the
bottom plots the black/thick lines are for the su(2) NJL model and the red/thin lines for the
chiral model I.}
\label{pressb}
\end{figure}

In a quark star, we must impose both $\beta$-equilibrium and charge neutrality. The relations between the chemical potentials of different particles are given by
\beq
\mu_d=\mu_u+\mu_e, \qquad\: \mu_e=\mu_\mu.
\eeq
In terms of the neutron and the  electron chemical potentials $\mu_n$ and $\mu_e$, one has
\beq
\mu_u=\frac{1}{3}\mu_n-\frac{2}{3}\mu_e, \: \qquad\: \mu_d=\frac{1}{3}\mu_n+\frac{1}{3}\mu_e.
\eeq
For the charge neutrality we impose
\beq
\rho_e+\rho_\mu=\frac{1}{3}\left(2\rho_u-\rho_d \right).
\eeq
\begin{figure}[ht]
\vspace{1.5cm}
\centering
\includegraphics[width=0.9\linewidth,angle=0]{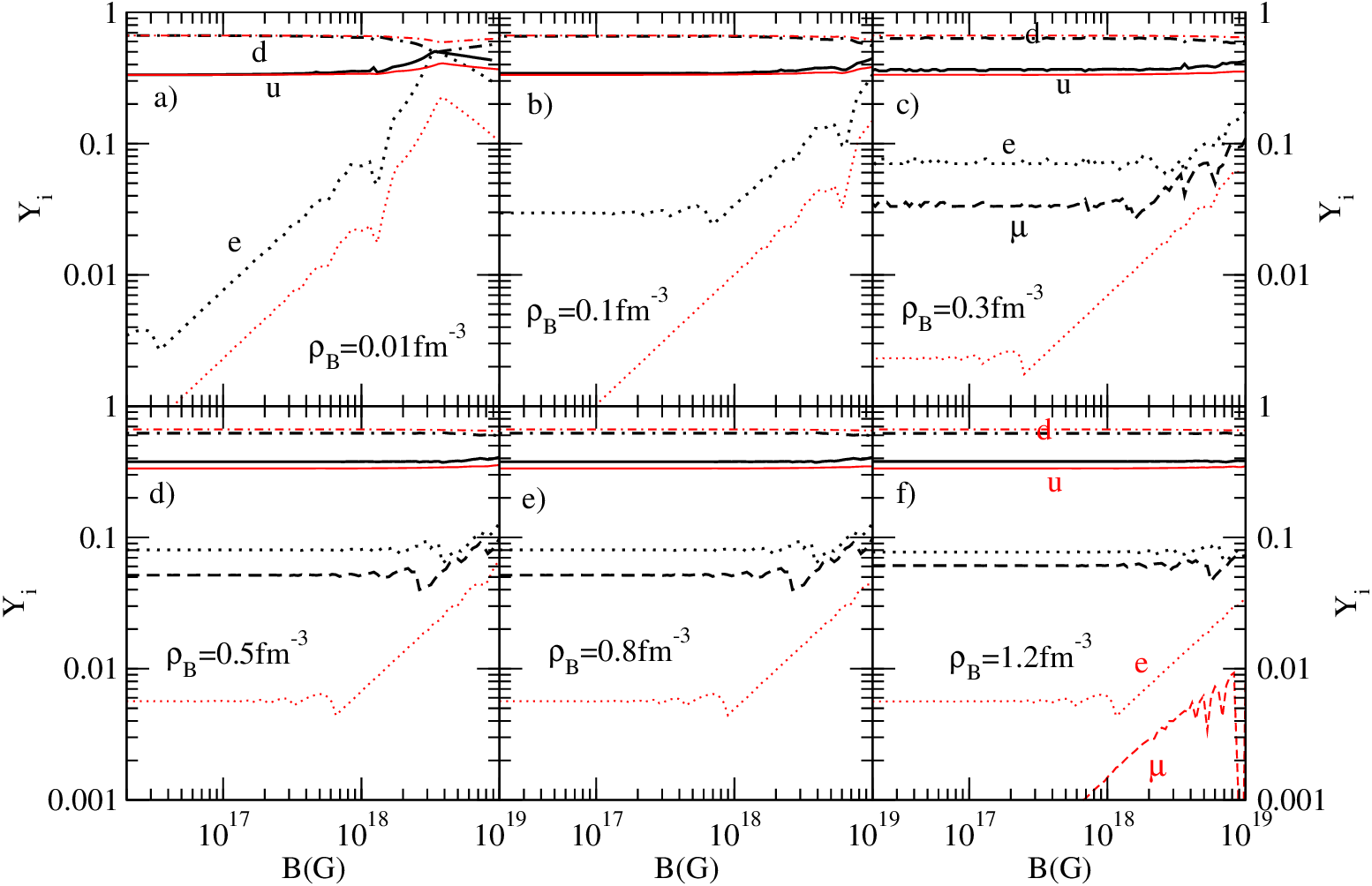}
\caption{(Color online) Quark and lepton fractions as a function of the magnetic field for asymmetric quark
matter, and for several values of the baryonic density ($\rho=0.001, \, 0.01,\, 0.3,\, 0.5, \,
0.8,\, 1.2$ fm$^{-3}$). The black/thick lines are  for the su(2) NJL model and red/thin lines for the chiral model.}
\label{frac}
\end{figure}

In Fig.~\ref{pressb}, we plot the pressure and the particle fraction as a
function of the baryon density obtained with the chiral model and the NJL model for different
values of the magnetic field intensity. For the chiral model, we will only consider the B
dependent vacuum effects through the renormalization of the pion decay constant, chiral model I. Since we will not go beyond $B=10^{19}$ G both prescriptions introduced in section \ref{sec:para} give similar results.

At the surface of the quark star, defined by a zero pressure,  the density is finite. The chiral
model predicts larger baryon densities at the surface and a softer EOS, for magnetic field below $10^{19}$ G. For larger fields the opposite occurs, {\it e.g. } NJL predicts larger densities at the surface. However,
according to the scalar
virial theorem~\cite{virial} the interior magnetic field strength could be as large as
$B \sim 1-3\times 10^{18}$ G so, in principle, fields stronger than the ones represented in Fig.~\ref{pressb} will not occur in the interior of compact stars.

Considering the particle fractions obtained within both models, it is seen that the NJL model
predicts a larger $u$-quark fraction, and consequently,  larger  electron and muon fractions.
At $B=10^{19}$ G the main effects due to the magnetic field  occur below $\rho=0.25$ fm$^{-3}$.
The irregularity of the curves is due to the filling of the Landau levels, which may give rise
to strong fluctuations on the particle fractions. There is a large increase
of the $u$-quark, and correspondingly of the electron fraction, with the increase of $B$. It is even observed that the $u$ quark fraction is 
larger than the $d$ quark fraction in a small range of densities. However, we should point out that according to the pressure plots the density at the surface of the star will be above 0.5 fm$^{-3}$. Therefore, the strong effects on the particle fractions below that density will not affect the star properties, except if the star has a crust as discussed in~\cite{glend95,reddy06}. In this case strong effects could occur in the star crust. The onset of the muon is sensitive to the field and reflects the
filling of the lepton Landau levels; larger muon fractions are attained  at high density due
to the larger $u$ quark fractions. For the chiral model the muon fraction is always below 0.001 in the absence of a magnetic field. This changes for a magnetic field stronger than 10$^{18}$ G for densities above 1 fm$^{-3}$ as seen in Fig.~\ref{pressb} and~\ref{frac}.

 In order to better understand the effect of the magnetic field  on the particle fractions we
plot in Fig.~\ref{frac} the particle fractions as a function of the magnetic field intensity
for six representative densities: the densities below 0.3 fm$^{-3}$ would only occur in the
crust of the star, in the case it exists,  $\sim$ 0.5 fm$^{-3}$ is the surface density, and
0.8 and 1.2  fm$^{-3}$ are baryon densities in the interior of the star. At the surface the
electron and muon fractions are larger for the NJL model. This has implications in the possible
existence of a crust~\cite{glend95}. A larger electron fraction will be able to support a larger
crust. In~\cite{menezes06} it was shown that NJL would predict a larger electron fraction at
the surface than the MIT bag model.  The chiral model seems to behave more like the MIT bag
model.
At the surface of a star with no crust,  the effect of the magnetic field starts to be
non-negligible for $B> 7\times 10^{17}$ G with a clear increase of the electron fraction in the
chiral model. This effect will occur for much smaller fields in a star with a crust as can be
seen in Fig.~\ref{frac}a). As a result, it is expected that  the structure of the crust will be
deeply influenced by the presence of a very strong magnetic field.

The effect of the magnetic field on the muon fraction within the chiral model is also clearly seen in this figure: for $\rho=1.2$ fm$^{-3}$ the muon fractions rises above 0.001 for fields a bit below 10$^{18}$ G.


\section{Conclusions}

In the present work we have compared the properties of quark matter under a strong magnetic
field described using three different models: the MIT bag model~\cite{bag} which describes
quark matter in a chiral restored phase,  the chiral model~\cite{kbk90,KRS84}, and the two flavor
NJL model~\cite{njl} both described by a chiral symmetric Lagrangian density. We have discussed
the effect of the magnetic field on the vacuum properties and how the parameters of the models
related to the vacuum should be chosen in order to take into account vacuum
corrections. However, these corrections are only important for very strong magnetic fields $B>
3\times 10^{19}$ G, which are not expected to be found in compact stars but could be formed
as short-lived magnetic fields in relativistic heavy ion collisions playing an important role
in possible experimental signatures of strong CP violation and the phenomenon of the chiral
magnetic effect~\cite{cp}. Estimations done in~\cite{skokov09} show that for the LHC energy it
could be possible to get $eB\sim 15\, m^2_\pi$ which corresponds to a field $B\sim 5\times
10^{19}$ G.

It was shown that if the schematic MIT quark model is used to describe quark matter under
strong magnetic fields the value of the bag pressure should be adjusted in order to account for
the magnetic field vacuum corrections. In the chiral model the vacuum corrections may be
taken into account by fitting the parameters of the model to a quark vacuum mass that includes
these corrections, chiral model I, or by including $B$ dependent vacuum corrections as
discussed in \cite{mizher10}, chiral model II. It was shown that for very strong fields, above
$5\times 10^{19}$ G, the chiral model I failed to include adequatly the $B$ dependent vaccum
correction. However, below those extreme magnetic fields chiral model I gave reasonable
results. Although chiral model II behaves like the NJL model there are some important
differences: its vacuum mass does not increase so fast as the one of the NJL when $B$
increases. As a result, in the chiral model II the energy per particle in the chiral symmetric broken phase becomes
smaller than the one of NJL for very large $B$ values. In the chiral symmetric phase the
effective bag pressure increases faster with $B$ in the NJL model.
 Therefore,  although the   bag pressure is smaller in the NJL for $B=0$,  at $B=10^{20}$ G, NJL and
the chiral model II have the same bag pressure.

Finally, we have applied the chiral model and the two flavor NJL model to the description of
stellar matter. It was shown that within the chiral model quark stars will have a larger baryon
density at the surface. This feature could be reversed for a strong magnetic field larger than
$3\times 10^{19}$. However, so strong magnetic fields are not expected to exist in compact
stars. Another difference is the larger (smaller) $u$ ($d$) quark fractions in the NJL
model. As a consequence the chiral model predicts much smaller lepton fractions. One of the
main effects of the magnetic field is to increase the $u$ quark fractions at low densities due
to their larger absolute charges,  and,
therefore, also the electron fractions. This could have strong effects on the strucutre of the 
crust of a quark star like the one predicted in~\cite{reddy06}.
  
\begin{acknowledgments}
We thank Jo\~ao da Provid\^encia for  valuable discussions.
A. R. specially acknowledges  helpful discussions with Prafulla K.
Panda at the beginning of this work.
This work was partially supported by FEDER and FCT (Portugal) under the projects
CERN/FP/109316/2009  and PTDC/FIS/113292/2009, and by Compstar, an ESF Research Networking
Programme.
\end{acknowledgments}

\end{document}